# Resolving Phonons in Superconductor $Bi_2Sr_2CaCu_2O_{8+\delta}$ at Sub-Unit-Cell Resolution

**Authors:** Xiaowen Zhang,[1,2]† Jiade Li,[1,2]† Xiaoyue Gao,[1,2] Ruochen Shi,[1,2] Bo Han,[1,2] Xiaomei Li,[3] Jinlong Du,[2] Jinsheng Wen,[4,5] Genda Gu,[6] Shichong Wang,[7] Wentao Zhang,[7]* Peng Gao[1,2,8]*.

**Affiliations:**

[1] International Center for Quantum Materials, School of Physics, Peking University; Beijing 100871, China.

[2] Electron Microscopy Laboratory, School of Physics, Peking University; Beijing 100871, China.

[3] School of Integrated Circuits, East China Normal University; Shanghai, 200241, China.

[4] National Laboratory of Solid State Microstructures and Department of Physics; Nanjing University, Nanjing 210093, China.

[5] Collaborative Innovation Center of Advanced Microstructures and Jiangsu Physical Science Research Center, Nanjing University; Nanjing 210093, China.

[6] Condensed Matter Physics and Materials Science Department, Brookhaven National Laboratory; Upton, New York 11973, USA.

[7] Beijing National Laboratory for Condensed Matter Physics, Institute of Physics, Chinese Academy of Sciences; Beijing 100190, China.

[8] Hefei National Laboratory; Hefei 230088, China.

†These authors contributed equally to this work.

*Corresponding author. Email: *wentaozhang@iphy.ac.cn; pgao@pku.edu.cn*

**Abstract**

The role of phonons in cuprates remains controversial, with their complex lattice structure complicating the investigation. Here, we identify phonon modes originating from charge reservoir and superconducting layers of $Bi_2Sr_2CaCu_2O_{8+\delta}$ using sub-unit-cell resolved electron energy loss spectroscopy in a scanning transmission electron microscope. We reveal several phonon modes exhibiting layer-specific localization: ~78 meV in-plane modes localized in the $CuO_2$ planes, ~42 meV hybrid (in-plane and out-of-plane) modes in the $CuO_2$ planes, and ~38 meV hybrid modes in the BiO layers. We also observe a periodic modulation of phonon frequencies induced by the superstructure, spatially correlating with the superconducting gap variation. Our findings offer new insights into the electron-phonon coupling in cuprates, fostering deeper exploration of the microscopic linkage between lattice dynamics and superconductivity.

**Main Text:**

The mechanism of high-temperature superconductivity in cuprates remains one of the most intriguing mysteries in condensed matter physics. While electron-phonon coupling is well-established as the driving force behind conventional superconductivity, its role in cuprates is still highly controversial (*1-7*). This debate largely centers on the intricate phonon spectra of cuprates (*8-11*), which render the contribution of specific phonon modes elusive and challenging to discern (*12*).

The unique structural complexity of cuprate superconductors gives rise to their complex phonon features—take the prototypical system $Bi_2Sr_2CaCu_2O_{8+\delta}$ (Bi-2212) as an example, whose crystal structure is composed of alternating superconducting $CuO_2$ planes and non-superconducting charge reservoir layers. This quasi-two-dimensional layered architecture gives rise to distinct local quasiparticle states in different functional layers (*13*). Additionally, although earlier studies have revealed phonon mode-selective coupling in Bi-2212 (*14*) and demonstrated its potential connection to superconducting electron pair condensation (*15*), the distinct roles of phonon modes in different functional layers remain unclear. Pinpointing these phonon modes may offer a viable approach to re-evaluating their differentiated contributions.

Additionally, the in-plane incommensurate supermodulation of the Bi-2212 lattice further exacerbates this complexity. This wave-like distortion, triggered by the mismatch in optimal bond lengths between the $CuO_2$ planes and charge reservoir layers, forms a superstructure with a period of approximately 2.6 nm (*16*). Intriguingly, the superconducting gap undergoes a same periodic modulation with this superstructure, strongly implying a potential connection to the superconducting pairing mechanism (*17*). Resolving the spatial changes of phonon modes induced by the superstructural modulation offers the potential for a quantitative evaluation of the extent to which phonons mediate superconductivity.

Overall, characterizing phonon modes at the sub-unit-cell scale, including interlayer differences in the out-of-plane direction and superstructural modulation in the in-plane direction, is essential for revealing the interplay between lattice dynamics and superconductivity. However, the traditional phonon detection techniques (*11, 18-21*) such as inelastic X-ray/neutron scattering (IXS/INS) and Raman/infrared spectroscopy cannot achieve nanoscale spatial resolution, only providing statistical average information from large area including numerous unit-cells or supermodulations. The recent advances in scanning transmission electron microscopy (STEM) and electron energy loss spectroscopy (EELS) have enabled atomic scale resolution to probe local phonons (*22-28*), providing new opportunity to reveal the spatial dependent phonons in cuprate superconductors.

Here, we report sub-unit-cell resolved phonon modes in Bi-2212 using scanning transmission electron microscopy with electron energy loss spectroscopy (STEM-EELS), enabling characterization of anisotropic lattice dynamics. Difference analysis of EEL spectra unveils three distinct phonon modes localized to the $CuO_2$ planes and charge reservoir layers. Real-space EELS mapping, combining with density functional theory (DFT) calculations, identify the pure in-plane phonon modes of $CuO_2$ planes, as well as the out-of-plane/in-plane hybrid phonon modes of $CuO_2$ planes and BiO layers. Critically, we directly observe a sinusoidal modulation of ($CuO_2$ planes) phonon frequencies induced by the superstructure, which aligns with the periodicity of the superconducting gap modulation, thereby establishing a direct link between lattice dynamics and superconducting gap in cuprates. These results provide new insights for understanding the complex electron-phonon coupling in cuprate high-temperature superconductors.

## Results

### Spatially Resolved Phonon Modes in Bi-2212

Figure 1A illustrates the crystal structure of Bi-2212, which is divided into two regions: (1) the Cu-O part (blue arrow), comprising the $CuO_2$ plane and the intervening Ca layer, and (2) the Bi-O part (orange arrow), consisting of the BiO and SrO layers. Due to the quasi-two-dimensional nature of Bi-2212, its lattice dynamics exhibit strong anisotropy with distinctly different in-plane (vibrations perpendicular to the *c*-axis) and out-of-plane (vibrations parallel to the *c*-axis) phonon modes. To identify the anisotropic phonon modes on different layers, we performed controlled off-axis geometry STEM-EELS measurements along the $[1\bar{1}0]$ zone axis (the gray plane in Fig. 1A).

In the off-axis geometry, the differential scattering cross-section of EEL spectra depends on the factor $|e_\lambda(k,q)\cdot q/q^2|^2$, where $e_\lambda(k,q)$ is the eigenvector of phonon mode $\lambda$ of the *k*th atom in the unit cell and $q$ is the momentum transfer collected by the EELS aperture (*26*) (green disks in Fig. 1B and Fig. 1E). This allows us to selectively probe phonon modes along specific momentum direction by controlling the off-axis direction (*28*). Figures 1C and 1F show the EEL spectra obtained from off-axis configurations in the $q_x$ (Fig. 1C, *i.e.* the in-plane) and $q_z$ (Fig. 1F, *i.e.* the out-of-plane) directions, respectively. Both configurations reveal significant differences in the EEL spectra between the Cu-O and Bi-O parts, confirming spatial sensitivity at the sub-unit-

cell level. These differences can be more easily identified in the mean EEL curves extracted from different parts in both off-axis configurations (Fig. 1D and Fig. 1G). It is important to note that the differences in phonon signals between the two spatial parts in the $q_x$ and $q_z$ are distinct (see Fig. S1 for details), reflecting the vibrational anisotropy of Bi-2212.

Although our raw EEL spectra have already demonstrated the ability to resolve phonon modes at the sub-unit-cell level, the delocalization of the vibrational signals in STEM-EELS measurements causes most of the phonon modes to spatially merge together (*22, 25*). To mitigate delocalization effects in STEM-EELS, we applied a differential processing method on the EEL spectra (see details in Materials and Methods), yielding refined the differential (Diff.) EEL spectra. Figures 2A and 2C show the Diff. EEL spectra of Bi-2212 along the $q_x$ and $q_z$ directions, respectively, which clearly distinguish between the Cu-O and Bi-O part. We extracted the mean Diff. EEL curves for each part to better illustrate the experimental findings (Figs. 2B and 2D). These spectra consistently show three prominent phonon features: F1, localized in the Bi-O part at approximately 38 meV; F2, in the Cu-O part at around 43 meV; F3, in the Cu-O part at about 78 meV. This indicates that F1 is associated with the vibrations of the Bi-O parts, while F2 and F3 originate from the Cu-O parts, which is consistent with the results from DFT calculations and spatially resolved EEL spectra simulations (Fig. S2, see details in Materials and Methods). Above results unequivocally demonstrate the capability of STEM-EELS to resolve phonon modes in cuprates at sub-unit-cell spatial and meV energy resolutions—a key advancement over traditional techniques.

**Anisotropic Vibrations in Real Space**

To obtain the vibrational properties of the phonon modes, we integrated the EELS signals within a specific frequency range and mapped the real-space distribution of phonon excitations (Fig. 3A and Fig. 3B for $q_x$ and $q_z$, respectively). In the F3 frequency range (76 – 80 meV), the real-space distribution is strongly localized in the Cu-O part for $q_x$ (Fig. 3A), but featureless for $q_z$ (Fig. 3B), indicating purely in-plane vibrations. In the F2 frequency range (42 – 46 meV), the real-space distribution is localized in the Cu-O part for both $q_x$ and $q_z$, indicating hybrid vibrations both in-plane and out-of-plane. Similarly, in the F1 frequency range (36 – 40 meV), the real-space distribution is localized in the Bi-O part for both $q_x$ and $q_z$, also indicating hybrid in-plane and out-of-plane vibrations. Furthermore, for the EEL spectra map along $q_x$ direction, while both the F2 and F3 frequency range maps exhibit higher intensity where the $CuO_2$ plane are located, their intensity ratio reaches maximums between the $CuO_2$ planes, as shown in Fig. S3. This spatial distribution of the intensity ratio suggests that the F2 phonon modes have more pronounced out-of-plane vibrations compared to the F3 phonon modes (*29*). The above results are confirmed by the calculated phonon dispersions with localization weights (Fig. S4B) and vibrational direction weights (Fig. S4C).

The real-space distribution of phonon-excitation intensity along different off-axis directions reflects the anisotropic vibrations of phonon modes in Bi-2212. To quantify the degree of anisotropy, we analyzed thermal ellipsoids from DFT calculations (Fig. 3C), defining the anisotropic ratio as $((U_{11}^2 + U_{22}^2) +1)/(U_{33}^2+1)$, where $U_{11}$, $U_{22}$ (in-plane) and $U_{33}$ (out-of-plane) are vibrational amplitudes. Larger deviations from 1 indicate stronger anisotropy. According to the thermal ellipsoids and the ratios in three frequency ranges, we find that the vibrations in the F3 range are localized on the oxygen atoms of the $CuO_2$ plane, and the anisotropic ratio (~ 3.7) is the furthest from 1. In the F2 and F1 ranges, the vibrations are localized on the oxygen atoms of the $CuO_2$ plane and BiO plane, with anisotropic ratios of ~ 0.4 and ~ 0.5, respectively. The anisotropy of phonon-mode vibrations between in-plane and out-of-plane directions leads to corresponding anisotropies in phonon-related properties, such as the electron-phonon coupling strength (*30*).

**Superstructure-Induced Phonon Modulation**

To investigate the modulation effects of the superstructure on phonon modes, we acquired the EEL spectra of the two-period superstructure, as shown by the gray box in Fig. 4A. The stacking of the EEL spectra along the in-plane direction (gray arrow in Fig. 4A) reveals periodic shifts of the EEL spectra (Fig. 4B). The peak positions of F2, and F3 also exhibit periodic behaviors as the superstructure modulation progresses. To more clearly demonstrate the influence of the superstructure on phonon modes, in Fig. 4C, we plotted the changes in the characteristic peaks of Bi-2212 with the superstructure phase Φ. The peak positions of F3 varies approximately as a cosine function with phase, the peak positions of F2 follows nearly a sine function, while the peak positions of F1 shows no significant change. As the phase Φ varies from 0° to 180°, the peaks' positions of F3 shift by approximately 2.4 meV (a relative variation of ~ 3%), while F2 exhibits a shift of ~ 2.1 meV (~ 5% change). HAADF imaging (Fig. 4D) reveals superstructure-driven lattice distortions: as the phase Φ advances from 0° to 180°, the in-plane distances shrink, leading to lattice compression in the plane. This results in an increase in in-plane phonon frequencies. Meanwhile, the out-of-plane distances expand, causing lattice stretching in the perpendicular direction, which corresponds to a decrease in out-of-plane phonon frequencies. Notably, the observed phonon frequency modulations (~ 3 – 5%) are approximately half the magnitude of the superconducting gap modulation (~ 10%) (*17*). Phonon frequency modulations and the superconducting gap modulation exhibit differences in phase and amplitude. These observations may contribute to understanding the connection between the superconducting gap and electron-phonon coupling in cuprates.

**Discussions and Conclusions**

In cuprates research, experimental investigations into the electron states have been extensively studied through both reciprocal-space (*e.g.* angle-resolved photoemission spectroscopy (ARPES)) (*31-33*) and real-space methods (*e.g.* scanning tunneling microscope (STM)) (*34-36*). These results have corroborated and complemented each other, enabling a comprehensive understanding of the electron states in cuprates. However, phonon studies have largely remained confined to reciprocal space (*11, 19-21*) due to limited spatial resolution. Our real-space investigation of phonon modes in Bi-2212 bridges this gap, offering new insights into their role in high-$T_c$ superconductivity. For example, the 'kink' feature in ARPES usually represents the strong electron-boson coupling strength (*37*). In previous studies, the 'kink' feature at 70–80 meV in the nodal states' region of the bulk Bi-2212 has triggered extensive controversy (*38*). Some researchers argue that this feature originates from the electron-phonon coupling (*39-41*), while others maintain that it stems from the electron-magnon (the collective excitation of the spin structure) coupling (*42-44*). Our experimental results provide the direct observation of a ~ 78 meV in-plane vibrational phonon mode localized on the $CuO_2$ planes (F3). Additionally, we observe two phonon modes (F2 and F1) near 40 meV, originating from the $CuO_2$ planes and the charge reservoir layers respectively. Their frequency scales closely match that of the 'kink' feature in the anti-nodal states' region of the bulk Bi-2212 (*45*). These observations provide new insights for re-examining the electron-boson coupling hierarchy in cuprates, particularly in disentangling phononic and magnetic contributions to the self-energy.

Furthermore, we uncovered how the superstructure, a periodic lattice distortion, modulates phonon modes in Bi-2212. The superstructure induces local variations in interatomic distances, leading to systematic softening or hardening of phonon modes. According to previous STM studies (*17*), the superconducting gap of Bi-2212 reaches its maximum value when the phase $\Phi = 0°$ or $360°$, while it attains the minimum value at $\Phi = 180°$. In our study, we discovered that: at $\Phi = 0°$, the out-of-plane vibrational frequency reaches its maximum value, and the in-plane vibrational frequency is at its minimum; conversely, at $\Phi = 180°$, the out-of-plane vibrational frequency drops to its minimum value, and the in-plane vibrational frequency rises to its maximum value. These behaviors provide valuable clues to evaluate the role of local electron-phonon coupling in cuprates on the superconductivity in future.

In conclusion, by means of sub-unit-cell resolved STEM-EELS experiments conducted in different momentum transfer directions, we disentangled the contributions of $CuO_2$ planes and charge reservoir layers to the phonon spectra, revealing their anisotropic properties. Our results linked the 78 meV mode to in-plane vibrations localized in the $CuO_2$ planes, offering new insights to the controversy about 'kink' features in ARPES. Moreover, we demonstrated how periodic lattice distortions modulate phonon states, providing a potential connection between local electron-phonon coupling and superconductivity. These findings provide new insights into understanding the role of local phonons and electron-phonon couplings in the superconducting mechanisms in copper-based high-temperature superconductors.

**Acknowledgments**

P.G. acknowledges the support from the New Cornerstone Science Foundation through the XPLORER PRIZE. We acknowledge Electron Microscopy Laboratory of Peking University for the use of electron microscopes, and the High-performance Computing Platform of Peking University for providing computational resources.

**Funding:** This work was supported by the National Natural Science Foundation of China (52125307 for P.G.; 12141404 for W.T.Z.; 12404192 for R.C.S.), the National Key R&D Program of China (2021YFA1400502 for P.G.) and the China Postdoctoral Science Foundation (Grant No. GZB20240028 for J.D.L.). The work at BNL was supported by the US Department of Energy, Office of Basic Energy Sciences, contract No. DOE-SC0012704.

**Author contributions:** X.W.Z. and J.D.L. contributed equally to this work. P.G. and W.T.Z. conceived the project; J.S.W. and G.D.G prepared the sample; X.W.Z. performed the STEM-EELS experiment and data analysis assisted by R.C.S., J.D.L. and X.Y.G. with the guidance of W.T.Z. and P.G.; X.W.Z. performed *ab initio* calculations; X.Y.G. prepared the TEM sample and acquired the atomic resolution HAADF images; X.W.Z., J.D.L., and P.G. wrote the manuscript; All the authors contributed to this work through useful discussion and/or comments to the manuscript.

**Competing interests:** The authors declare that they have no competing interests.

**Data and materials availability:** All data needed to evaluate the conclusions in the paper are available in the main text or the supplementary materials. Additional data related to this paper may be requested from the authors.

**Supplementary Materials:** Materials and Methods; Figs. S1 to S7.

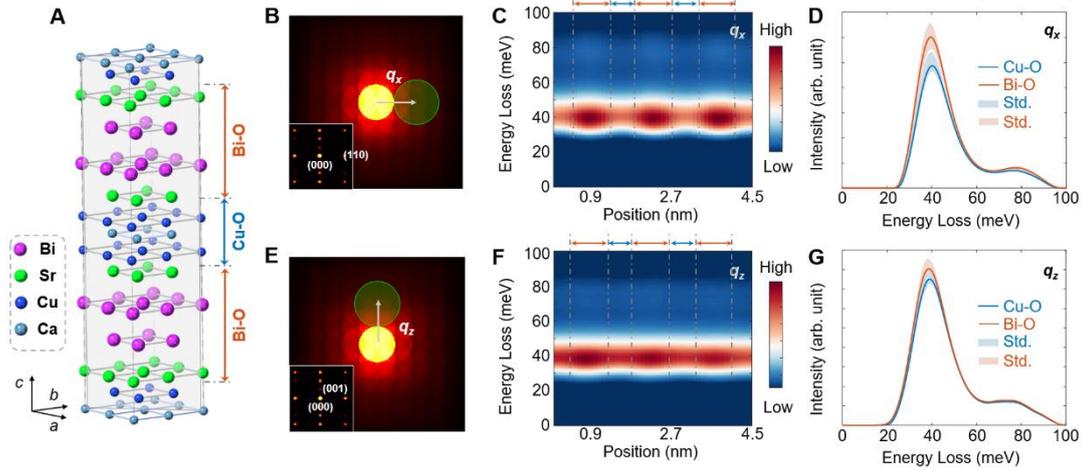

**Fig. 1 Experimental settings and EEL spectra of Bi-2212.** (**A**) Schematic of Bi-2212 structure. All oxygen atoms are hidden for clarity, with the minor deformation along the *b*-axis omitted. The Cu-O part is labeled with the vertical blue arrow, while the Bi-O part is labeled with the vertical orange arrow. The faint gray plane indicates the projected plane in STEM-EELS experiments. (**B**, **E**) Projected momentum spaces. Green circles indicate EELS aperture configurations with the net momentum transfer directions perpendicular to the *c*-axis (labeled as $q_x$) and parallel to the *a*-axis (labeled as $q_z$), respectively. Lower-left insets: diffraction patterns of Bi-2212. (**C**, **F**) Line profile of EEL spectra of Bi-2212 along $q_x$ and $q_z$, respectively. Blue (orange) arrow marked the location of the Cu-O (Bi-O) part of Bi-2212. (**D**, **G**) EEL spectra extracted from different regions of (**C**, **F**), respectively. Error bands (labeled as Std.) represent the standard deviation of the intensities of the EEL spectra over different spatial positions.
11

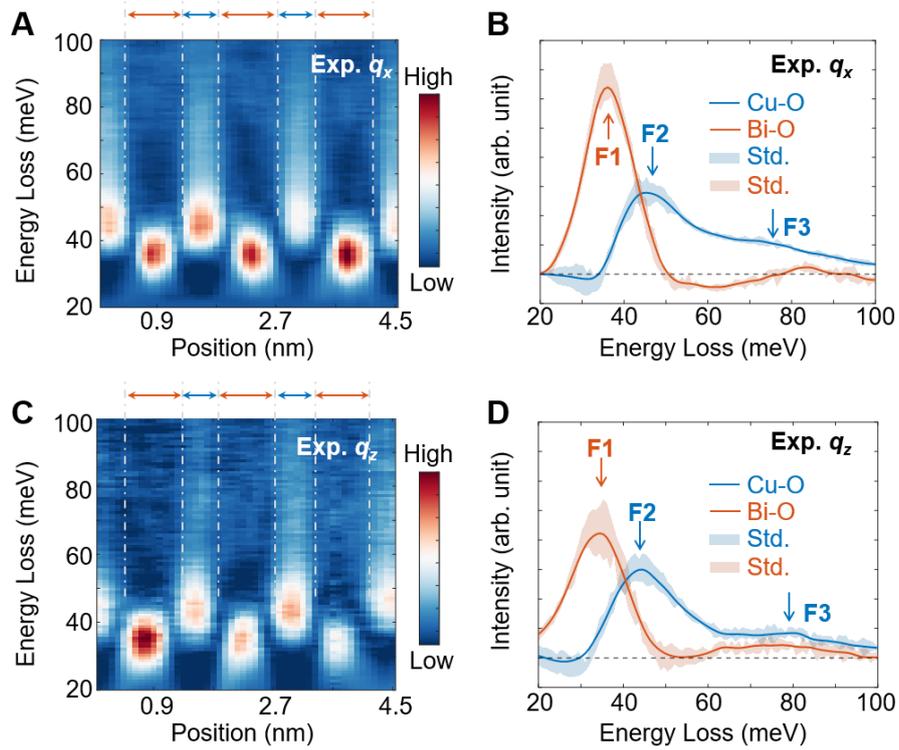

**Fig. 2 Differential EEL spectra of Bi-2212.** (**A**, **C**) Line profile of the Differential (Diff.) EEL spectra derived from Fig. 1(**C**, **F**), respectively. Blue (orange) arrow marked the location of the Cu-O (Bi-O) part of Bi-2212. (**B**, **D**) The Diff. EEL spectra extracted from the different part of (**A**, **C**), respectively. F1-F3 labels three prominent features. Error bands (labeled as Std. region) represent the standard deviation of the intensities of the Diff. EEL spectra over different spatial positions.

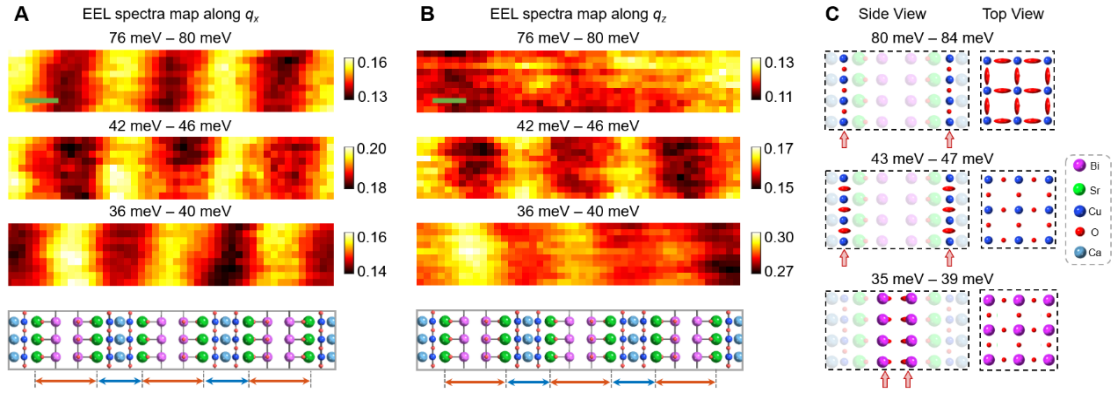

**Fig. 3 Real-space phonon excitation intensity maps of Bi-2212.** (**A**) Real-space distribution of the phonon-excitation intensity for $q_x$ in the frequency range 76 – 80 meV (top), 42 – 46 meV (middle) and 36 – 40 meV (bottom) respectively. (**B**) Results in the $q_z$ direction corresponding to (**A**). Scale bar in (**A**) and (**B**) is 0.5 nm. The atomic structure corresponding to the scanned region is displayed below spatial maps. (**C**) Projected thermal ellipsoids derived from DFT calculations show atomic mean-squared displacements in the frequency ranges 80 – 84 meV (top), 43 – 47 meV (middle) and 35 – 39 meV (bottom). Red arrows indicate the vibrating atomic layers, while atoms in non-vibrating layers are rendered transparent. For visual clarity, all displacement magnitudes are exaggerated.

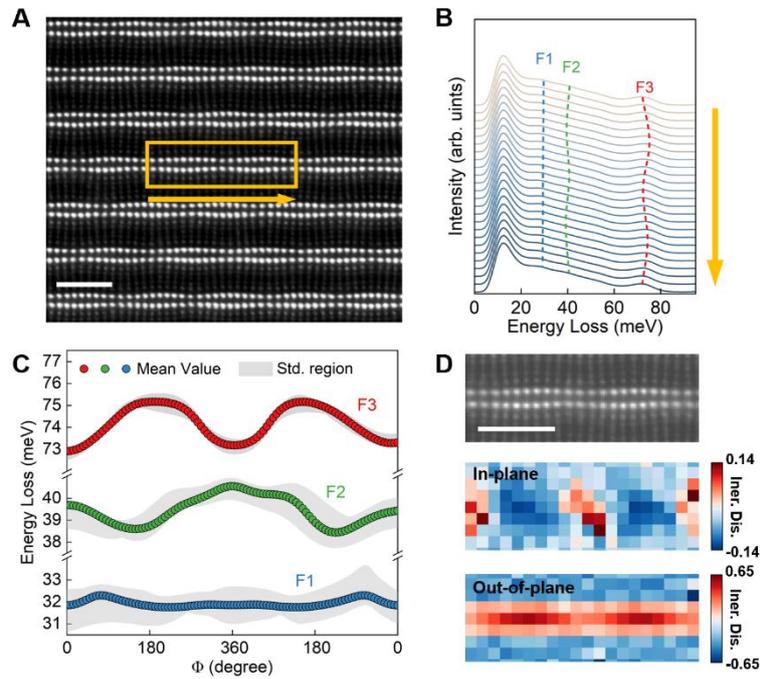

**Fig. 4 Superstructure modulation of phonon modes.** (**A**) HAADF image of Bi-2212. The scanning region and scanning direction are labeled by the yellow box and arrows, respectively. Scale bar, 2 nm. (**B**) Stacked EEL spectra along the scanning direction from top to bottom. Blue, green, and red dashed curves mark the peak positions of F1, F2, and F3, respectively. (**C**) Variation of the peak positions of F1 (blue points), F2 (green points), and F3 (red points). The light gray shadows represent the standard deviation of the fitted peak positions. (**D**) Top: HAADF image of the superstructure of Bi-2212. Scale bar, 2 nm. Middle and bottom: Ratio of interatomic distance (abbreviated as Inter. Dis.) variation of in-plane and out-of-plane directions extracted from the HAADF image.

14